\documentclass[aps,prd,preprint,eqsecnum,floats,epsf,superscriptaddress,nofootinbib]{revtex4}
\usepackage{epsfig}
\usepackage{graphicx}
\usepackage{extarrows}
\usepackage{subfigure}

\begin{document}
\def\be{\begin{eqnarray}}
\def\en{\end{eqnarray}}
\def\non{\nonumber}
\def\ov{\overline}
\def\la{\langle}
\def\ra{\rangle}
\def\B{{\cal B}}
\def\pr{{\sl Phys. Rev.}~}
\def\prl{{\sl Phys. Rev. Lett.}~}
\def\pl{{\sl Phys. Lett.}~}
\def\np{{\sl Nucl. Phys.}~}
\def\zp{{\sl Z. Phys.}~}
\def\up{\uparrow}
\def\dw{\downarrow}
\def\lsim{ {\ \lower-1.2pt\vbox{\hbox{\rlap{$<$}\lower5pt\vbox{\hbox{$\sim$}
}}}\ } }
\def\gsim{ {\ \lower-1.2pt\vbox{\hbox{\rlap{$>$}\lower5pt\vbox{\hbox{$\sim$}
}}}\ } }

\font\el=cmbx10 scaled \magstep2{\obeylines\hfill April, 2022}

\vskip 1.5 cm

\centerline{\large\bf Heavy-Flavor-Conserving Hadronic Weak Decays }
\centerline{\large\bf of Charmed and Bottom Baryons}

\bigskip
\bigskip
\centerline{\bf Hai-Yang Cheng$^1$, Fanrong Xu$^2$}
\medskip
\centerline{$^1$Institute of Physics, Academia Sinica}
\centerline{Taipei, Taiwan 115, Republic of China}
\medskip
\medskip
\centerline{$^2$Department of Physics, Jinan University}
\centerline{Guangzhou 510632, People's Republic of China}
\bigskip
\bigskip
\centerline{\bf Abstract}
\bigskip
\small
Three decades ago, heavy-flavor-conserving (HFC) weak decays of heavy baryons such as $\Xi_Q\to\Lambda_Q\pi$ and $\Omega_Q\to\Xi_Q\pi$ for $Q=c,b$ had been studied
within the framework that incorporates both heavy-quark and chiral symmetries. It was pointed out that if the heavy quark in the HFC process behaves as a spectator, then the $P$-wave amplitude of $\B_{\bar 3}\to \B_{\bar 3}+\pi$ with $\B_{\bar 3}$ being an antitriplet heavy baryon will vanish in the heavy quark limit. Indeed, this is the case for $\Xi_b\to\Lambda_b \pi$ decays. For $\Xi_c\to\Lambda_c\pi$ decays, they receive  additional nonspectator contributions arising from the $W$-exchange diagrams through the $cs\to dc$ transition. Spectator and nonspectator $W$-exchange contributions to the $S$-wave amplitude of $\Xi_c\to\Lambda_c\pi$ are destructive, rendering the $S$-wave contribution even smaller.
However, the nonspectator effect on the $P$-wave amplitude was overlooked in all the previous model calculations until a very recent investigation within the framework of a constituent quark model in which the parity-conserving pole terms were found to be dominant in  $\Xi_c\to\Lambda_c\pi$ decays. Since the pion produced in the HFC process is soft, we apply current algebra to study both
$S$- and $P$-wave amplitudes and employ the bag and diquark models to estimate the matrix elements of four-quark operators. We confirm that $\Xi_c\to\Lambda_c\pi$ decays are indeed dominated by the parity-conserving transition induced from  nonspectator $W$-exchange and that they receive largest contributions from the intermediate $\Sigma_c$ pole terms.
We also show that the $S$-wave of $\Omega_b\to \Xi_b \pi$ decays vanishes in the heavy quark limit, while $\Omega_c\to \Xi_c\pi$ receive additional $W$-exchange contributions via $cs\to dc$ transition. The $P$-wave contribution to $\Omega_c\to \Xi_c\pi$ is enhanced by the ${\Xi'}_c$ pole, though it is not so dramatic as in the case of $\Xi_c\to\Lambda_c\pi$.
The asymmetry parameter $\alpha$ is found to be positive, of order $0.70$ and $0.74$ for
$\Xi_c^0\to \Lambda_c^+\pi^-$ and $\Xi_c^+\to \Lambda_c^+\pi^0$, respectively.
The predicted branching fraction is of order $5\times 10^{-4}$ for $\Omega_c^0\to\Xi_c^+\pi^-$ and $3\times 10^{-4}$ for $\Omega_c^0\to\Xi_c^0\pi^0$ both with the asymmetry parameter close to $-1$.

\pagebreak

\section{Introduction}
Thirty years ago, heavy-flavor-conserving (HFC) weak decays of heavy baryons such as the singly
Cabibbo-suppressed decays $\Xi_Q\to\Lambda_Q\pi$ and $\Omega_Q\to\Xi_Q\pi$ for $Q=c,b$ were first studied in Ref. \cite{Cheng:HFC} within the framework that incorporates both heavy-quark and chiral symmetries \cite{Yan,Wise}.
Since then, these HFC decays   have been studied in \cite{Sinha:HFC,Voloshin00,Voloshin14,Faller,Gronau:2015jgh,Gronau:HFC,Cheng:HFC2016,%
Voloshin:2019,Groote:2021pxt,Niu:2021qcc}.
Unlike hadronic weak decays of heavy mesons, a rigorous and reliable approach for describing the nonleptonic decays of heavy baryons does not exist. Nevertheless, there is a special class of weak decays of heavy baryons that
can be studied in a more trustworthy way, namely, HFC
nonleptonic decays. In general, the evaluation of nonfactorizable contributions to heavy baryon decays $\B_i\to \B_f+P$ decays often relies on the pole model in which one considers the contributions from all possible intermediate states. In common practice, the $S$-wave amplitude is dominated by the low-lying $1/2^-$
resonances while the $P$-wave one governed by the ground-state
$1/2^+$ poles. However, the calculation of $S$-wave
amplitudes is far more difficult than the $P$-wave ones owing to the
troublesome negative-parity baryon resonances which are not well
understood in the quark model. It is well known that in the soft-pion limit, the PV (parity-violating) amplitude is reduced
to a simple commutator term expressed in terms of PC (parity-conserving)
matrix elements. Since the emitted light mesons are soft in nonleptonic HFC decays, the nonfactorizable $S$-wave amplitude can be evaluated reliably using current algebra.
Therefore, one of the great advantages of this special class of weak decays
is that the evaluation of the PV
$S$-wave amplitude does not require the information
of the negative-parity $1/2^-$ poles.

As for the $P$-wave amplitude, if only
the light quarks inside the heavy baryon participate in weak
interactions and the heavy quark behaves as a ``spectator", then the $P$-wave amplitude will vanish in the heavy quark limit. The argument goes as follows.
In the heavy quark limit, the diquark of the antitriplet heavy baryon $\B_{\bar 3}$ is a scalar one with $J^P=0^+$. If only the light quarks involve in weak interactions, the weak diquark transition will be $0^+\to 0^+ + 0^-$ for  $\B_{\bar 3}\to \B_{\bar 3}+P$. Based on the conservation of angular momentum, it is easily seen that the parity-conserving $P$-wave amplitude vanishes. Hence, the $P$-wave amplitude of the $\B_{\bar 3}\to \B_{\bar 3}+P$ decay vanishes in the heavy quark limit provided that the heavy quark acts as a spectator \cite{Cheng:HFC}. Indeed, this is the case for $\Xi_b\to\Lambda_b\pi$ decays. This is the second major advantage for the study of HFC decays.

However, both charm-flavor-conserving decays $\Xi_c\to \Lambda_c^+\pi$ and $\Omega_c\to \Xi_c\pi$ receive additional contributions from the $W$-exchange diagrams via $cs\to dc$ transition. Although the charm flavor is still conserved in those modes,  $P$-wave  will not diminish in the heavy quark limit because of the nonspectator $W$-exchange contributions.

LHCb has measured the $b$-flavor-conserving and strangeness-changing weak decay $\Xi_b^-\to\Lambda_b^0\pi^-$ \cite{LHCb:HFCb}. The relative rate was measured to be
\be
{f_{\Xi_b^-}\over f_{\Lambda_b^0}} \B(\Xi_b^-\to\Lambda_b^0\pi^-)=(5.7\pm1.8^{+0.8}_{-0.9})\times 10^{-4},
\en
where $f_{\Xi_b^-}$ and $f_{\Lambda_b^0}$ are $b\to \Xi_b^-$ and $b\to\Lambda_b^0$ fragmentation fractions, respectively. Assuming $f_{\Xi_b^-}/ f_{\Lambda_b^0}$ in the range between 0.1 and 0.3, based on the measured production rates of other strange particles relative to their non-strange counterparts \cite{LHCb:HFCb}, the branching fraction of
$\Xi_b^-\to\Lambda_b^0\pi^-$ will lie in the range
\be \label{eq:BFXib LHCB}
\B(\Xi_b^-\to\Lambda_b^0\pi^-)=(0.57\pm0.21)\sim(0.19\pm0.07)\%.
\en
The charm-flavor-conserving decay $\Xi_c^0\to\Lambda_c^+\pi^-$ first advocated and studied in 1992 \cite{Cheng:HFC} was finally measured by the LHCb in 2021 with a surprisingly large branching fraction of $(0.55\pm0.02\pm0.18)\%$ \cite{LHCb:HFC}.

As we shall see below in Table \ref{tab:ComparisonBR}, all the previous studies of charm-flavor-conserving weak decays before 2021 have neglected $P$-wave contributions. The largest branching fraction $\B(\Xi_c^0\to\Lambda_c^+\pi^-)=(1.34\pm0.53)\times 10^{-3}$ was obtained in Ref. \cite{Gronau:HFC}, assuming a constructive interference between
the $W$-exchange $S$-wave amplitudes induced by $su\to ud$ and $cs\to dc$ transitions. However, as pointed out in \cite{Cheng:HFC2016}, the interference should be destructive. Thus the predicted branching fraction will be of order $(1\sim 3)\times 10^{-4}$ which is too small compared to the LHCb measurement. This issue was finally resolved last year. It was pointed out in Refs. \cite{Niu:2021qcc,Groote:2021pxt} that the $P$-wave amplitude induced through $cs\to dc$ $W$-exchange was overlooked in all the previous model calculations. It turns out that owing to the small mass difference between $\Xi_c$ and the intermediate $\Sigma_c$ pole,
PC amplitudes
are one to two orders of magnitude larger than the PV ones \cite{Niu:2021qcc}. That is,  $\Xi_c\to\Lambda_c^+\pi$ receives largest contributions from the $\Sigma_c$ pole terms.

The study of HFC decays in Ref. \cite{Niu:2021qcc} was performed in the non-relativistic constituent quark model. The calculation is rather complicated and tedious. For example, the $S$-wave is evaluated by taking into account various negative-parity heavy baryon resonances. In this work we shall rely on the fact that the pion's momentum is 104 and 116 MeV, respectively, in $\Xi_b\to\Lambda_b^0\pi$ and $\Xi_c\to\Lambda_c^+\pi$ decays, it is thus legitimate to apply for the soft-pion theorem to  both $S$- and $P$-wave amplitudes. We then employ the MIT bag model and the diquark model to evaluate the matrix elements of four-quark operators. The goal is to see if the HFC $\Xi_c\to\Lambda_c\pi$ decays are indeed dominated by the PC terms.

In the heavy quark limit, the weak diquark transition responsible for HFC $\Omega_Q\to \Xi_Q\pi$ decays is $1^+\to 0^++ 0^-$. As a consequence, the factorizable contribution to
$\Omega_b\to \Xi_b\pi$ is purely a $P$-wave, in contrast to the case of $\Xi_b\to\Lambda_b\pi$. In the charm sector, nonspectator $W$-exchange contributes to $\Omega_c\to\Xi_c\pi$ and the $\Xi'_c$ pole will give rise to an enhancement to the $P$-wave amplitude, though its effect is not as dramatic as in the case of $\Xi_c\to\Lambda_c\pi$.

This work is organized as follows. In Secs. II and III we study $S$- and $P$-wave amplitudes, respectively, for HFC decays of heavy baryons. Some model calculations are presented in Sec. IV followed by some discussions in Sec. V. Sec. VI gives our conclusion.

\section{Parity-violating $S$-wave amplitudes}
The relevant effective $\Delta S=1$ weak Hamiltonian for HFC weak decays reads
\be  \label{eq:H}
H_{\rm eff}={G_F\over \sqrt{2}}V^*_{ud}V_{us}(c_1O_1+c_2O_2)+h.c.,
\en
and the four-quark operators are given by
\be
O_1=(\bar du)(\bar us), \qquad O_2=(\bar ds)(\bar uu),
\en
with $(\bar q_1q_2)\equiv \bar q_1\gamma_\mu(1-\gamma_5)q_2$.
The Wilson coefficients to the leading order are given by
$c_1=1.336$ and $c_2=-0.621$ obtained at the scale $\mu=m_c$ with $\Lambda^{(4)}_{\ov {\rm MS}}=325$ MeV
and $c_1=1.139$ and $c_2=-0.307$ at the scale $\mu=\bar m_b(m_b)$ with $\Lambda^{(5)}_{\ov {\rm MS}}=225$ MeV \cite{Buchalla}.
As pointed out in \cite{Voloshin00}, there is an additional ``nonspectator" $W$-exchange contribution to heavy-flavor-conserving decays of charmed baryons governed by
\be \label{eq:Hc}
H_{\rm eff}^{(c)}={G_F\over \sqrt{2}}V^*_{cd}V_{cs}(c_1\tilde O_1+c_2\tilde O_2)+h.c.,
\en
with $\tilde O_1=(\bar dc)(\bar cs)$ and $\tilde O_2=(\bar cc)(\bar d s)$.

The general amplitude for $\B_i\to \B_f+P$ reads
\begin{eqnarray} \label{eq:Amp}
M(\B_i\to \B_f+P)=i\bar u_f(A-B\gamma_5)u_i,
\end{eqnarray}
where $A$ and $B$ are the $S$- and $P$-wave amplitudes, respectively.
Each amplitude consists of factorizable and nonfactorizable ones
\be
A=A^{\rm fact}+A^{\rm nf}, \qquad B=B^{\rm fact}+B^{\rm nf}.
\en
While the factorizable amplitude vanishes in the soft meson limit, the nonfactorizable one is not. We shall study the PV $S$-wave amplitude first.

The factorizable contributions are given by
\be
\la \pi^-\Lambda_c^+|H_{\rm eff}|\Xi_c^0\ra^{\rm fac} &=& {G_F\over\sqrt{2}}V_{ud}^*V_{us}\,a_1\la \pi^-|(\bar du)|0\ra\la \Lambda_c^+|(\bar us)|\Xi_c^0\ra, \non \\
\la \pi^0\Lambda_c^+|H_{\rm eff}|\Xi_c^+\ra^{\rm fac} &=& {G_F\over\sqrt{2}}V_{ud}^*V_{us}\,a_2\la \pi^0|(\bar uu)|0\ra\la \Lambda_c^+|(\bar ds)|\Xi_c^+\ra,
\en
with
\be \label{eq:a1a2}
a_1=c_1+{c_2\over N_c}, \qquad a_2=c_2+{c_1\over N_c},
\en
In terms of the form factors defined by
\be
\la \Lambda_c^+|(\bar u s)|\Xi_c^0\ra &=& \bar u_{\Lambda_c}\Big[f_1^{\Lambda_c\Xi_c}(q^2)\gamma_\mu+f_2^{\Lambda_c\Xi_c}(q^2)i\sigma_{\mu\nu}q^\nu
+f_3^{\Lambda_c\Xi_c}(q^2)q_\mu \non \\
&& -g_1^{\Lambda_c\Xi_c}(q^2)\gamma_\mu\gamma_5-g_2^{\Lambda_c\Xi_c}(q^2)i\sigma_{\mu\nu}q^\nu\gamma_5
-g_3^{\Lambda_c\Xi_c}(q^2)q_\mu\gamma_5\Big]u_{\Xi_c},
\en
and $\la \pi^-(q)|A_\mu|0\ra=if_\pi q_\mu$, \footnote{We follow the PDG convention $\la 0|A_\mu|P(q)\ra=if_Pq_\mu$ \cite{PDG}. A charge conjugate of $\la 0|\bar u\gamma_\mu\gamma_5 d|\pi^+(q)\ra=if_\pi q_\mu$ leads to $\la \pi^-(-q)|\bar d\gamma_\mu\gamma_5 u|0\ra=-if_\pi q_\mu$ and hence $\la \pi^-(q)|A_\mu|0\ra=if_\pi q_\mu$.
}
we obtain
\be \label{eq:Facamp}
\la \pi^-\Lambda_c^+|H_{\rm eff}|\Xi_c^0\ra^{\rm fac} &=& -i{G_F\over\sqrt{2}}V_{ud}^*V_{us}\, a_1f_\pi \bar u_{\Lambda_c}\Big[(m_{\Xi_c}-m_{\Lambda_c})f_1^{\Lambda_c\Xi_c}(m_\pi^2) \non \\
&& +(m_{\Xi_c}+m_{\Lambda_c})g_1^{\Lambda_c\Xi_c}(m_\pi^2)\gamma_5\Big]u_{\Xi_c}.
\en
Hence,
\be
A(\Xi_c^0\to\Lambda_c^+\pi^-)^{\rm fac} &=& -{G_F\over\sqrt{2}}V_{ud}^*V_{us}\,a_1f_\pi (m_{\Xi_c}-m_{\Lambda_c})f_1^{\Lambda_c^+\Xi_c^0}(m_\pi^2), \non \\
A(\Xi_c^+\to\Lambda_c^+\pi^0)^{\rm fac} &=& -{G_F\over 2}V_{ud}^*V_{us}\,a_2f_\pi (m_{\Xi_c}-m_{\Lambda_c})f_1^{\Lambda_c^+\Xi_c^+}(m_\pi^2).
\en

Nonfactorizable contributions arise from the $W$-exchange diagrams to be shown below. One popular approach for evaluating the $W$-exchange ampitudes
is to consider the contributions from all possible intermediate states, including resonances and continuum states. In practice, one usually focuses on the most important poles such as the low-lying $1/2^+$ and $1/2^-$ states. More specifically, the $S$-wave amplitude is dominated by the low-lying $1/2^-$ resonances and the $P$-wave one governed by the ground-state  poles.  The nonfactorizable $S$- and $P$-wave amplitudes for the process $\B_i\to \B_f+M$ are then given by  \cite{Cheng:1992}
\be \label{eq:amppole}
A^{\rm pole} &=& -\sum_{\B_n^*(1/2^-)}\left[ {g_{_{\B_f\B_{n^*}M}b_{n^*i}}\over m_i-m_{n^*}}+ {b_{fn^*}g_{_{\B_{n^*}\B_iM}}\over m_f-m_{n^*}}\right]+\cdots, \non \\
B^{\rm pole} &=& -\sum_{\B_n}\left[ {g_{_{\B_f\B_nM}} a_{ni}\over m_i-m_{n}}+ {a_{fn}g_{_{\B_{n}\B_iM}}\over m_f-m_{n}}\right]+\cdots,
\en
respectively. Ellipses in the above equation denote other pole contributions which are negligible for our purposes, and the baryon-baryon matrix elements are defined by \cite{Cheng:1992}
\be
\la \B_i|H_{\rm eff}|\B_j\ra=\bar u_i(a_{ij}+b_{ij}\gamma_5)u_j, \qquad
\la \B_i^*(1/2^-)|H_{\rm eff}^{\rm pv}|\B_i\ra = b_{i^*j}\bar u_i u_j.
\en

In Ref. \cite{Niu:2021qcc}, the nonfactorizable $S$-wave amplitude was evaluated by considering the first orbital excited sextet states $\Sigma_c^{0,+}$, ${\Xi'}_c^+$ and the antitriplet states $\Lambda_c^+$ and $\Xi_c^+$ with the quantum number $1/2^-$ for $\Xi_c\to\Lambda_c^+\pi$. However, the calculation is inevitably rather complicated and the uncertainties are large because the excited negative-parity states are far from being well established. In view of the soft pion nature in HFC decays, the tedious $S$-wave calculation can be greatly simplified. Applying the Goldberger-Treiman relation for the strong coupling $\B'\B P$ and its generalization for $\B^*\B P$
\be \label{eq:GT}
g_{_{\B'\B P^a}}= {\sqrt{2}\over f_{P^a}}(m_{\B'}+m_\B)g^A_{\B'\B}, \qquad
g_{_{\B^*\B P^a}}= {\sqrt{2}\over f_{P^a}}(m_{\B^*}-m_\B)g^A_{\B^*\B},
\en
to Eq. (\ref{eq:amppole}), then in the soft-meson limit, the intermediate excited $1/2^-$ states can be summed up and reduced to a commutator term (see e.g. the derivation in Ref.
\cite{CKX}); that is, $A^{\rm pole}\to A^{\rm com}$ in the soft-meson limit with
\begin{equation}
A^{\rm{com}}=-\frac{\sqrt{2}}{f_{P^a}}\la \mathcal{B}_f|[Q_5^a, {\cal H}_{\rm{eff}}^{\rm{pv}}]|\mathcal{B}_i\ra
= -\frac{\sqrt{2}}{f_{P^a}}\la \mathcal{B}_f|[Q^a, {\cal H}_{\rm{eff}}^{\rm{pc}}]|\mathcal{B}_i\ra\label{eq:Apole}
\end{equation}
and
\begin{equation}
Q^a=\int d^3x \bar q\gamma^0\frac{\lambda^a}{2}q,\qquad
Q^a_5=\int d^3x \bar q\gamma^0\gamma_5\frac{\lambda^a}{2}q.
\end{equation}
The above expression for $A^{\rm com}$ is precisely the well known
soft-pion theorem in the current-algebra approach. That is, in the
soft-pseudoscalar-meson limit, the PV amplitude is reduced
to a simple commutator term expressed in terms of PC
matrix elements.

Applying the soft pion theorem, we have the following nonfactorizable $S$-wave amplitude for $\Xi_c^0\to\Lambda_c^+\pi^-$
\begin{eqnarray} \label{eq:Anf}
A(\Xi_c^0\to\Lambda_c^+\pi^-)^{\rm nf} &=& -{1\over f_\pi}\la \Lambda_c^+|\,[ I_+, {\cal H}_{\rm eff}+ {\cal H}^{(c)}_{\rm eff}]\,|\Xi_c^0\ra \non \\
& =&  {1\over f_\pi}\la \Lambda_c^+|{\cal H}_{\rm eff}|\Xi_c^+\ra + {1\over f_\pi}\la \Lambda_c^+|{\cal H}^{(c)}_{\rm eff}|\Xi_c^+\ra, \\
&\equiv & A^{\rm nf}_{su\to ud}+ A^{\rm nf}_{cs\to cd} \non
\end{eqnarray}
where $I_+$ is the isopsin ladder operator with $I_+|d\ra=|u\ra$.
The above equation describes  the $W$-exchange contributions through the  $su\to ud$ and $cs\to dc$ transitions, respectively. It is straightforward to show that
\be \label{eq:AmpHFC_2}
\left. \begin{array}{c}  A^{\rm nf}_{su\to ud} \\
A^{\rm nf}_{cs\to cd} \end{array} \right\} = {G_F\over 2\sqrt{2} f_\pi}V_{ud}^*V_{us}(c_1-c_2) \left\{ \begin{array}{c} X \\ -Y, \end{array} \right.
\en
with
\be \label{eq:X&Y}
X &\equiv& \la \Lambda_c^+|(\bar du)(\bar us)-(\bar uu)(\bar ds)|\Xi_c^+\ra , \non \\
Y &\equiv& \la \Lambda_c^+|(\bar dc)(\bar sc)-(\bar cc)(\bar ds)|\Xi_c^+\ra.
\en
To derive Eq. (\ref{eq:AmpHFC_2}), we have employed the fact that the operator $O_1\pm O_2$ is symmetric (antisymmetric) in color indices, so that only $O_1-O_2$ contributes to the baryon-baryon matrix element.
The minus sign in front of $Y$ in Eq. (\ref{eq:AmpHFC_2}) comes the the relation for the CKM matrix elements $V_{cd}^*V_{cs}= -V_{ud}^*V_{us}$ to a very good approximation.
The final expression \cite{Cheng:HFC2016}
\be \label{eq:AmpHFCXic0}
A(\Xi_c^0\to\Lambda_c^+\pi^-) &=& A^{\rm fac}+A^{\rm nf}_{su \to ud} + A^{\rm nf}_{sc\to cd}  \\
&=& {G_F\over\sqrt{2} f_\pi}V_{ud}^*V_{us}\left[ -a_1f_\pi^2(m_{\Xi_c}-m_{\Lambda_c})f_1^{\Lambda_c^+\Xi_c^0}(m_\pi^2)+{1\over 2}(c_1-c_2)\left(X-Y\right)\right]  \non
\en
is quite general and model independent. The remaining task is to evaluate the matrix elements $X$ and $Y$.
For given flavor wave functions of $\Lambda_c$ and $\Xi_c$ (see Refs. \cite{CKX,Zou:2019kzq} for our convention), the relative signs between various terms are fixed.

For completeness, the other $S$-wave amplitudes of HFC decays of $\Xi_c^+$ and $\Xi_b^{-,0}$ read
\be \label{eq:AmpHFC}
A(\Xi_c^+\to\Lambda_c^+\pi^0) &=& {G_F\over 2 f_\pi}V_{ud}^*V_{us}\left[ -a_2f_\pi^2(m_{\Xi_c}-m_{\Lambda_c})f_1^{\Lambda_c^+\Xi_c^+}(m_\pi^2) + {1\over 2}(c_1-c_2)\left(X-Y\right)\right], \non \\
A(\Xi_b^-\to\Lambda_b^0\pi^-) &=& {G_F\over\sqrt{2} f_\pi}V_{ud}^*V_{us}\left[-a_1f_\pi^2(m_{\Xi_b}-m_{\Lambda_b})f_1^{\Lambda_b^0\Xi_b^-}(m_\pi^2)+ {1\over 2}(c_1-c_2)X\right],  \\
A(\Xi_b^0\to\Lambda_b^0\pi^0) &=& {G_F\over 2 f_\pi}V_{ud}^*V_{us}\left[-a_2f_\pi^2(m_{\Xi_b}-m_{\Lambda_b})f_1^{\Lambda_b^0\Xi_b^0}(m_\pi^2)+ {1\over 2}(c_1-c_2)X\right]. \non
\en
Note that $c_{1,2}$ and $a_{1,2}$ are evaluated at the scale  $\mu=m_c$ and $\bar m(m_b)$
for $\Xi_c$ and $\Xi_b$, respectively.

For $\Omega_Q$ with $Q=c,b$, its diquark is an axial-vector one with $J^P=1^+$. Therefore, if the heavy quark $Q$ behaves as a spectator, the weak diquark transition of the HFC decay $\Omega_Q\to \Xi'_Q\pi$ will be $1^+\to 1^++ 0^-$. Then both $S$- and $P$-wave transitions receive factorizable contributions as pointed out in \cite{Cheng:HFC}. However, $\Omega_Q\to \Xi'_Q\pi$ are not kinematically allowed for both $Q=c$ and $b$. For $\Omega_Q\to \Xi_Q\pi$ decays, the diquark transition is $1^+\to 0^+ + 0^-$. This implies that contrary to $\Xi_Q\to\Lambda_Q\pi$ decays with the diquark weak transition $0^+\to 0^+ + 0^-$, the $S$-wave vanishes in the heavy quark limit provided that the heavy quark does not
participate in weak interactions. To see this explicitly, we write down the PV amplitude
\be
A(\Omega_c^0\to\Xi_c^+\pi^-)
&=& -{G_F\over\sqrt{2} }V_{ud}^*V_{us} a_1f_\pi(m_{\Omega_c}-m_{\Xi_c})f_1^{\Xi_c^+\Omega_c^0}(m_\pi^2) - {1\over f_\pi}\la \Xi_c^0|{\cal H}_{\rm eff}+ {\cal H}_{\rm eff}^{(c)}|\Omega_c^0\ra, \\
A(\Omega_c^0\to\Xi_c^0\pi^0)
&=& -{G_F\over 2}V_{ud}^*V_{us} a_2f_\pi(m_{\Omega_c}-m_{\Xi_c})f_1^{\Xi_c^0\Omega_c^0}(m_\pi^2) +{1\over \sqrt{2}f_\pi}\la \Xi_c^0|{\cal H}_{\rm eff}+ {\cal H}_{\rm eff}^{(c)}|\Omega_c^0\ra.  \non
\en
In the bag model, the form factor $f_1^{\Xi_c^+\Omega_c^0}$ ($f_1^{\Xi_c^0\Omega_c^0}$)  is proportional to $\la \Xi_c^+|b^\dagger_u b_s|\Omega_c^0\ra$ ($\la \Xi_c^0|b^\dagger_d b_s|\Omega_c^0\ra$), which vanishes after applying the charmed baryon flavor wave functions given in Ref. \cite{CKX}.
As shown in Ref. \cite{Cheng:HFC}, the combined heavy quark and chiral symmetries severely restrict the weak transitions allowed. It turns out that $\B_{\bar 3}-\B_6$ weak transition via the weak Hamiltonian ${\cal H}_{\rm eff}$ is prohibited in the heavy quark limit,  namely,
\be \label{eq:3-6transition}
\la \B_{\bar 3}|{\cal H}_{\rm eff}|\B_6\ra=0,
\en
where $\B_6$ denotes the sextet heavy baryon.
Hence,
\be \label{eq:SwaveOMegac}
A(\Omega_c^0\to\Xi_c^+\pi^-)= -{1\over f_\pi}a_{\Xi_c^0\Omega_c^0}, \qquad
A(\Omega_c^0\to\Xi_c^0\pi^0)=  {1\over \sqrt{2}f_\pi}a_{\Xi_c^0\Omega_c^0},
\en
with $a_{\Xi_c^0\Omega_c^0}=\la \Xi_c^0|{\cal H}_{\rm eff}^{(c)}|\Omega_c^0\ra$. Obviously,
the $S$-wave contribution vanishes in the case of $\Omega_b^-\to\Xi_b \pi$ decays.

\section{Parity-conserving $P$-wave amplitudes}
We next turn to the PC amplitude. The factorizable $P$-wave reads from Eq. (\ref{eq:Facamp}) to be
\be
B(\Xi_c^0\to \Lambda_c^+\pi^-)^{\rm fac} =
{G_F\over\sqrt{2}}V_{ud}^*V_{us}\,a_1 f_\pi (m_{\Xi_c}+m_{\Lambda_c})g_1^{\Lambda_c\Xi_c}(m_\pi^2).
\en
The form factor $g_1^{\Lambda_c\Xi_c}$ vanishes in the heavy quark limit, as inferred from Eq. (3.26) of Ref. \cite{Yan}.

\begin{figure}[t]
\begin{center}
\includegraphics[width=0.70\textwidth]{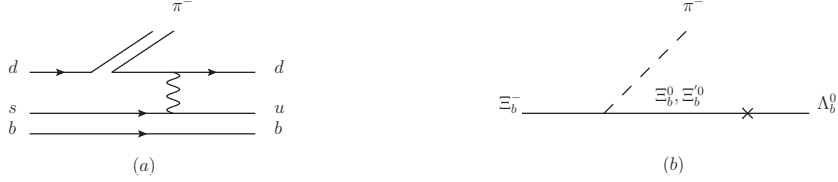}
\vspace{0.cm}
\caption{The $W$-exchange diagram for the HFC decay $\Xi_b^-\to \Lambda_b^0\pi^-$ and the corresponding pole diagram.} \label{fig:Xib-}
\end{center}
\end{figure}

\begin{figure}[t]
\begin{center}
\includegraphics[width=0.90\textwidth]{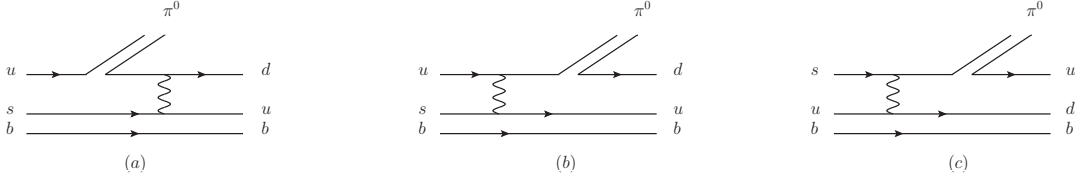}
\vspace{0.cm}
\caption{$W$-exchange diagrams for the HFC decay $\Xi_b^0\to \Lambda_b^0\pi^0$.} \label{fig:Xib0}
\end{center}
\end{figure}

As for the  nonfactorizable $P$-wave amplitude,  we notice that the diquark of $\B_{\bar 3}$ is a scalar one with $J^P=0^+$ in the heavy quark limit.
It was first pointed out in \cite{Cheng:HFC} that if the heavy quark behaves as a spectator, the weak diquark transition $0^+\to 0^+ + 0^-$ for  $\B_{\bar 3}\to \B_{\bar 3}+P$ is not allowed for the $P$-wave amplitude owing to the conservation of angular momentum. Hence, $P$-wave vanishes in the symmetry limit. To see this in the realistic calculation,
we apply the first Goldberger-Treiman relation in Eq. (\ref{eq:GT}) to Eq. (\ref{eq:amppole}) to get
\begin{equation}
B^{\rm{pole}}=-\frac{\sqrt{2}}{f_{P^a}}\sum_{\mathcal{B}_n}\left[ g^A_{\mathcal{B}_f \mathcal{B}_n}\frac{m_f+m_n}{m_i-m_n}a_{ni}
+a_{fn}\frac{m_i + m_n}{m_f-m_n} g_{\mathcal{B}_n \mathcal{B}_i}^A\right].
\label{eq:Bpole}
\end{equation}
Taking the decay $\Xi_b^-\to \Lambda_b^0\pi^-$ as an illustration,
we see from Fig. \ref{fig:Xib-}(b) that the relevant poles $\B_n$ are $\Xi_b^0$ and ${\Xi'}_b^0$. Likewise, the intermediate poles are $\Sigma_b^0,\Lambda_b^0,\Xi_b^0$ and ${\Xi'}_b^0$ for the decay $\Xi_b^0\to \Lambda_b^0\pi^0$  (see Fig. \ref{fig:Xib0}). Hence,
\be \label{eq:PwaveHFC}
B(\Xi_b^-\to \Lambda_b^0\pi^-)^{\rm pole} &=&
-{1\over f_\pi}\left(
 a_{\Lambda_b^0 \Xi_b^0}{m_{\Xi_b^-}+m_{\Xi_b^0}\over m_{\Lambda_b^0}-m_{\Xi_b^0}}g_{\Xi_b^0\Xi_b^-}^{A(\pi^-)}
 +  a_{\Lambda_b^0 {\Xi'}_b^0}{m_{\Xi_b^-}+m_{{\Xi'}_b^0}\over m_{\Lambda_b^0}-m_{{\Xi'}_b^0}}g_{{\Xi'}_b^0\Xi_b^-}^{A(\pi^-)}\right), \non  \\
B(\Xi_b^0\to \Lambda_b^0\pi^0)^{\rm pole} &=& -{\sqrt{2}\over f_\pi}\left(
g_{\Lambda_b^0\Sigma_b^0}^{A(\pi^0)}{m_{\Lambda_b^0}+m_{\Sigma_b^0}\over m_{\Xi_b^0}-m_{\Sigma_b^0}}a_{\Sigma_b^0 \Xi_b^0 } +
g_{\Lambda_b^0\Lambda_b^0}^{A(\pi^0)}{2 m_{\Lambda_b^0}\over m_{\Xi_b^0}-m_{\Lambda_b^0}}a_{ \Lambda_b^0 \Xi_b^0} \right.   \\
&&\quad + \left. a_{\Lambda_b^0 \Xi_b^0}{2m_{\Xi_b^0}\over m_{\Lambda_b^0}-m_{\Xi_b^0}}g_{\Xi_b^0\Xi_b^0}^{A(\pi^0)}+ a_{\Lambda_b^0 {\Xi'}_b^0}{m_{\Xi_b^0}+m_{{\Xi'}_b^0}\over m_{\Lambda_b^0}-m_{{\Xi'}_b^0}}g_{{\Xi'}_b^0\Xi_b^0}^{A(\pi^0)} \right), \non
\en
with
\be
a_{\Sigma_b^0 \Xi_b^0 }=\la \Sigma_b^0|{\cal H}_{\rm eff}| \Xi_b^0\ra, \qquad
a_{ \Lambda_b^0 \Xi_b^0}=\la \Lambda_b^0|{\cal H}_{\rm eff}| \Xi_b^0\ra, \qquad a_{\Lambda_b^0 {\Xi'}_b^{0}}= \la \Lambda_b^0|{\cal H}_{\rm eff}| {\Xi'}_b^{0}\ra.
\en

\begin{figure}[t]
\begin{center}
\includegraphics[width=0.70\textwidth]{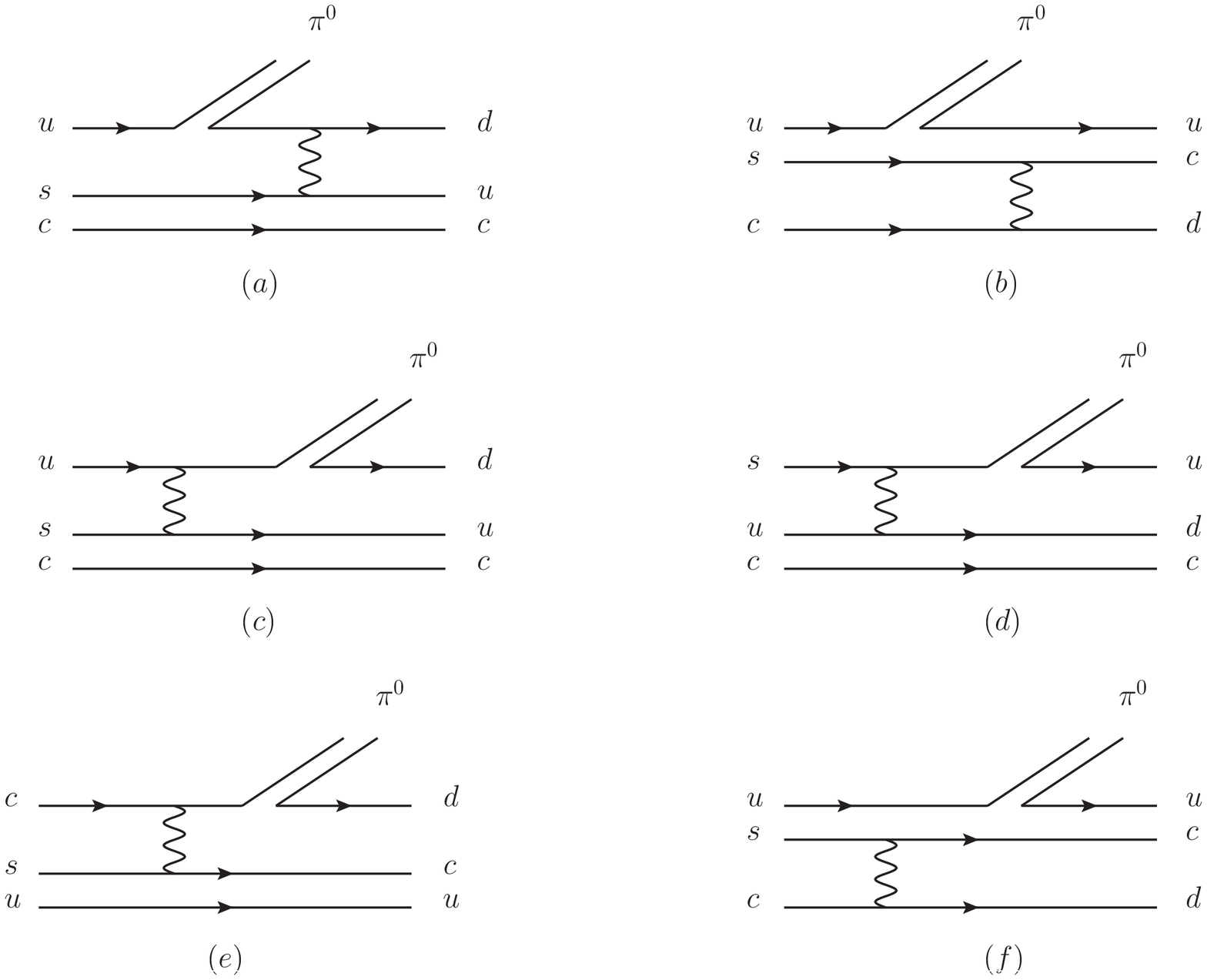}
\vspace{0.cm}
\caption{$W$-exchange diagrams for the HFC decay $\Xi_c^+\to \Lambda_c^+\pi^0$.} \label{fig:Xic+}
\end{center}
\end{figure}

\begin{figure}[t]
\begin{center}
\includegraphics[width=0.80\textwidth]{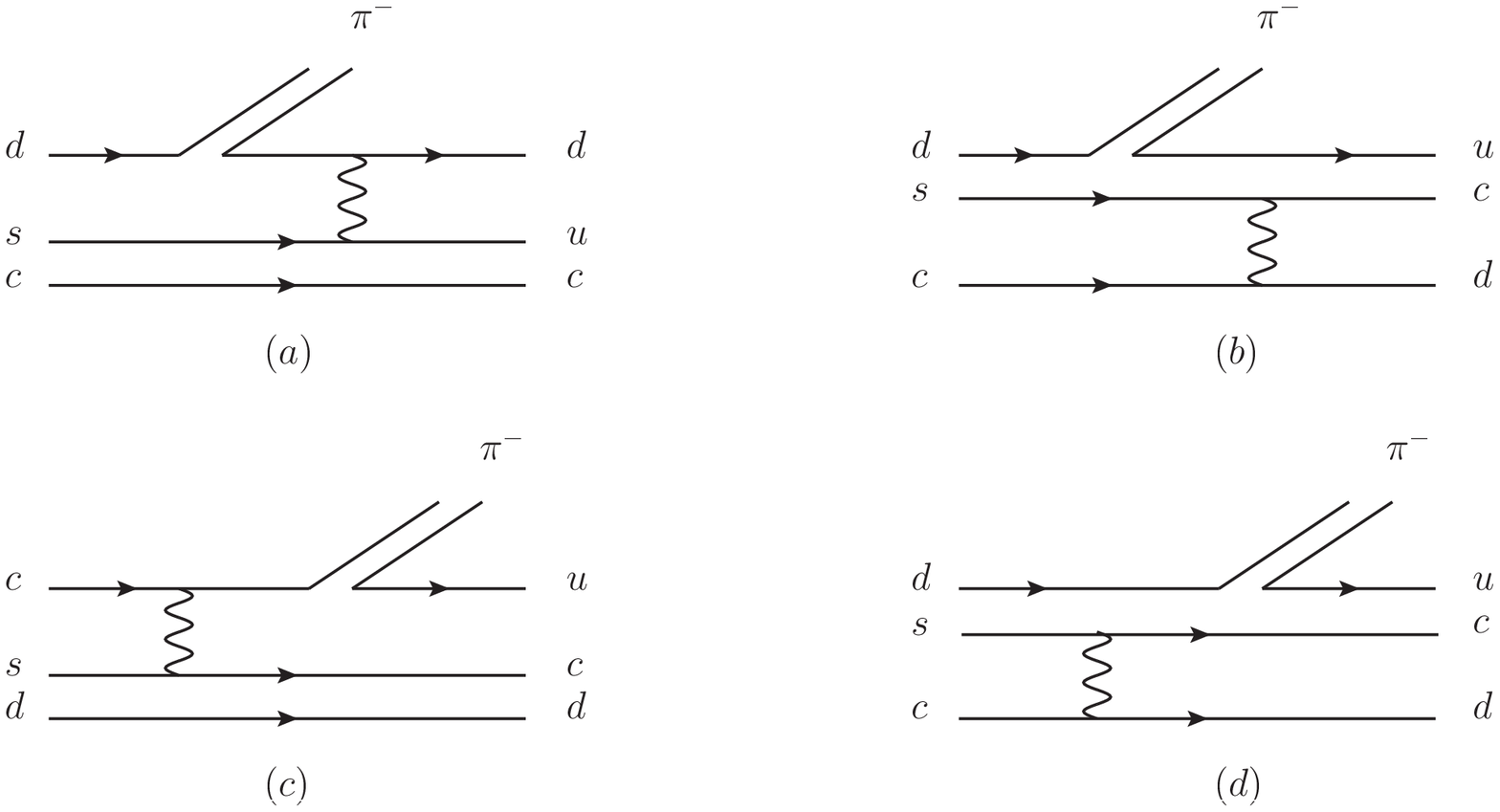}
\vspace{0.cm}
\caption{$W$-exchange diagrams for the HFC decay $\Xi_c^0\to \Lambda_c^+\pi^-$.} \label{fig:Xic0}
\end{center}
\end{figure}

The matrix elements $a_{\Sigma_b^0 \Xi_b^0}$ and $a_{\Lambda_b^0 {\Xi'}_b^{0}}$ vanish because of Eq. (\ref{eq:3-6transition}). Moreover, the axial-vector form factors $g_{\Xi_b^0\Xi_b^-}^{A(\pi^-)}$, $g_{\Xi_b^0\Xi_b^0}^{A(\pi^0)}$ and $g_{\Lambda_b^0\Lambda_b^0}^{A(\pi^0)}$ also vanish in the limit of heavy quark symmetry.
It had been shown in Ref. \cite{Yan} that the chiral Lagrangian Tr$(\bar \B_{\bar 3}\gamma_\mu \gamma_5 A^\mu \B_{\bar 3})$ vanishes in the heavy quark limit within the framework in which heavy quark symmetry and chiral symemtry are synthesized. This is a statement that the pion is emitted from the light quark and the transition of the diquark $0^+\to 0^++\pi$ is not permitted as it does not conserve parity. Hence, the ${\B_{\bar 3}\B_{\bar 3}\pi}$ coupling vanishes and so does the form factor $g_{\B_{\bar 3}\B_{\bar 3}}^A$ in the limit of heavy quark symmetry.
This symmetry  is also respected in the quark model because the spin operator at the weak vertex is prohibited.

Consequently, there is no any $P$-wave contributions to the HFC decays of $\Xi_b$. By the same token, nonfactorizable $P$-wave amplitudes in $\Xi_c\to\Lambda_c^+\pi$ decays receive zero contributions from the $W$-exchange through $su\to ud$ transition.
Referring to Figs. \ref{fig:Xic+} and \ref{fig:Xic0}, this means that Figs. \ref{fig:Xic+}(a), \ref{fig:Xic+}(c), \ref{fig:Xic+}(d) and \ref{fig:Xic0}(a) do not  contribute  to the $P$-wave amplitudes.
Nevertheless, as stressed in Refs. \cite{Niu:2021qcc,Groote:2021pxt}, the non-spectator  $W$-exchange through the $cs\to dc$ process will contribute to the PC amplitude. Explicitly,
\be \label{eq:Pwavenet}
B(\Xi_c^0\to \Lambda_c^+\pi^-)^{\rm pole} &=& - {1\over f_\pi}\left(
g_{\Lambda_c^+\Sigma_c^0}^{A(\pi^-)}{m_{\Lambda_c^+}+m_{\Sigma_c^0}\over m_{\Xi_c^0}-m_{\Sigma_c^0}}a_{\Sigma_c^0 \Xi_c^0 }
+  a_{\Lambda_c^+ {\Xi'}_c^+}{m_{\Xi_c^0}+m_{{\Xi'}_c^+}\over m_{\Lambda_c^+}-m_{{\Xi'}_c^+}}g_{{\Xi'}_c^+\Xi_c^0}^{A(\pi^-)}\right),  \non \\
B(\Xi_c^+\to \Lambda_c^+\pi^0)^{\rm pole} &=& - {\sqrt{2}\over f_\pi}\left(
g_{\Lambda_c^+\Sigma_c^+}^{A(\pi^0)}{m_{\Lambda_c^+}+m_{\Sigma_c^+}\over m_{\Xi_c^+}-m_{\Sigma_c^+}}a_{\Sigma_c^+ \Xi_c^+ }
+ a_{\Lambda_c^+ {\Xi'}_c^+}{m_{\Xi_c^+}+m_{{\Xi'}_c^+}\over m_{\Lambda_c^+}-m_{{\Xi'}_c^+}}g_{{\Xi'}_c^+\Xi_c^+}^{A(\pi^0)} \right),
\en
with
\be
a_{\Sigma_c^{0(+)} \Xi_c^{0(+)} }=\la \Sigma_c^{0(+)}|{\cal H}^{(c)}_{\rm eff}| \Xi_c^{0(+)}\ra, \qquad
 a_{\Lambda_c^+ {\Xi'}_c^+}= \la \Lambda_c^+ |{\cal H}^{(c)}_{\rm eff}|{\Xi'}_c^+ \ra.
\en
Since the masses of $\Xi_c$ and $\Sigma_c$ are very close, of order 16 MeV, it is obvious that the small mass difference between $\Xi_c$ and the $\Sigma_c$ pole gives rise to $(m_{\Lambda_c^+}+m_{\Sigma_c^+})/(m_{\Xi_c^+}-m_{\Sigma_c^+})=315$ and
$(m_{\Lambda_c^+}+m_{\Sigma_c^0})/(m_{\Xi_c^0}-m_{\Sigma_c^0})=276$,
which in turn lead to a strong enhancement of the $P$-wave pole amplitude \cite{Niu:2021qcc,Groote:2021pxt}.

\begin{figure}[t]
\begin{center}
\includegraphics[width=0.80\textwidth]{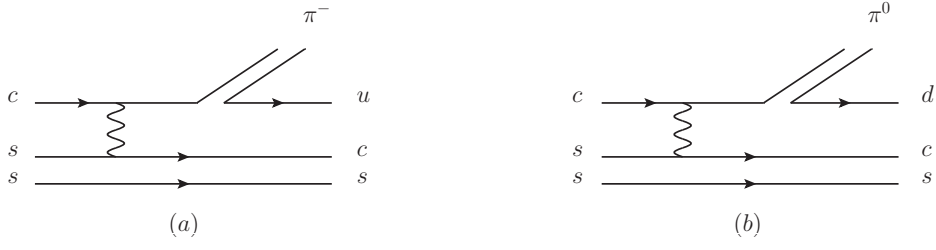}
\vspace{0.cm}
\caption{$W$-exchange diagrams for the HFC decays: (a) $\Omega_c^0\to \Xi_c^+\pi^-$ and (b) $\Omega_c^0\to \Xi_c^0\pi^0$.} \label{fig:Omegac}
\end{center}
\end{figure}

As noticed in passing, Contrary to $\Xi_Q\to\Lambda_Q\pi$ decays, $\Omega_Q\to \Xi_Q\pi$ do receive factorizable $P$-wave contributions as they are allowed by the diquark transition $1^+\to 0^+ +0^-$. Moreover, $\Omega_c\to \Xi_c\pi$ receive additional pole contributions from Fig. \ref{fig:Omegac}. Explicitly,
\be \label{eq:PwaveOmegaQ}
B(\Omega_c^0\to \Xi_c^+\pi^-) &=& {G_F\over \sqrt{2}}V_{ud}^*V_{us} \,a_1f_\pi(m_{\Omega_c}+m_{\Xi_c})g_1^{\Xi_c^+\Omega_c^0}(m_\pi^2) -
{1\over f_\pi}
g_{\Xi_c^+{\Xi'}_c^0}^{A(\pi^-)}\,{m_{\Xi_c^+}+m_{{\Xi'}_c^0}\over m_{\Omega_c^0}-m_{{\Xi'}_c^0}}a_{{\Xi'}_c^0 \Omega_c^0 },
 \non \\
B(\Omega_c^0\to \Xi_c^0\pi^0) &=& {G_F\over 2}V_{ud}^*V_{us} \,a_2f_\pi(m_{\Omega_c}+m_{\Xi_c})g_1^{\Xi_c^0\Omega_c^0}(m_\pi^2)
- {\sqrt{2}\over f_\pi}
g_{\Xi_c^0{\Xi'}_c^0}^{A(\pi^0)}\,{m_{\Xi_c^0}+m_{{\Xi'}_c^0}\over m_{\Omega_c^0}-m_{{\Xi'}_c^0}}a_{{\Xi'}_c^0 \Omega_c^0 }, \non \\
B(\Omega_b^-\to \Xi_b^0\pi^-) &=& {G_F\over \sqrt{2}}V_{ud}^*V_{us} \,a_1f_\pi(m_{\Omega_b}+m_{\Xi_b})g_1^{\Xi_b^0\Omega_b^-}(m_\pi^2), \\
B(\Omega_b^-\to \Xi_b^-\pi^0) &=& {G_F\over 2}V_{ud}^*V_{us} \,a_2f_\pi(m_{\Omega_b}+m_{\Xi_b})g_1^{\Xi_b^-\Omega_b^-}(m_\pi^2), \non
\en
with $a_{{\Xi'}_c^0 \Omega_c^0 }=\la {\Xi'}_c^0|{\cal H}^{(c)}_{\rm eff}|\Omega_c^0\ra$,
where we have dropped the contributions from the $\Xi_c$ pole as the form factor $g_{\Xi_c{\Xi}_c}^{A(\pi)}$ vanishes.

\section{Model Calculations}
To evaluate the nonfactorizable amplitudes we need to know the baryon matrix elements and the axial-vector form factor at $q^2=0$, namely, $g^A_{\B'\B}$.
There are seven four-quark matrix elements need to be evaluated, namely, $\la \Lambda_c^+|{\cal H}_{\rm eff}|\Xi_c^+\ra$, $\la \Lambda_c^+|{\cal H}^{(c)}_{\rm eff}|\Xi_c^+\ra$ (or $X$ and $Y$ defined in Eq. (\ref{eq:X&Y})), $a_{\Sigma_c^{0(+)} \Xi_c^{0(+)}}$, $a_{\Lambda_c^+ {\Xi'}_c^+}$, $a_{\Xi_c^0\Omega_c^0}$ and $a_{{\Xi'}_c^0\Omega_c^0}$.  We first consider the evaluation based on the MIT bag model \cite{MIT} (for details, see Appendices in Refs. \cite{Cheng:HFC2016,CKX,Zou:2019kzq}). We obtain
\be \label{eq:XY}
X =  32\pi X_2, \qquad  Y=8\pi (Y_1+Y_2),
\en
and
\be \label{eq:m.e.Xic}
a_{\Sigma_c^{+} \Xi_c^{+} }={1\over \sqrt{2}}a_{\Sigma_c^{0} \Xi_c^{0} }= -a_{\Lambda_c^+ {\Xi'}_c^+}={G_F\over 2\sqrt{2} }V_{cd}^*V_{cs}(c_1-c_2) {2\over \sqrt{3}}(-Y_1+3Y_2)(4\pi),
\en
with the bag integrals
\be  \label{eq:X2Y1Y2}
X_2 &=& \int_0^R r^2 dr(u_du_u+v_dv_u)(u_su_u+v_sv_u), \non \\
Y_1 &=& \int_0^R r^2 dr(u_dv_c-v_du_c)(u_sv_c-v_su_c), \\
Y_2 &=& \int_0^R r^2 dr(u_du_c+v_dv_c)(u_su_c+v_sv_c), \non
\en
where $u(r)$ and  $v(r)$ are the large and small components of the quark wave function, respectively. For the matrix elements relevant to $\Omega_c^0$, we find
\be \label{eq:meofOmegaQ}
a_{\Xi_c^0\Omega_c^0} &=& {G_F\over 2\sqrt{2}}V_{cd}^*V_{cs}(c_1-c_2) 2\sqrt{2\over 3}(Y_1-3Y_2)(4\pi), \non \\
a_{{\Xi'}_c^0\Omega_c^0} &=& - {G_F\over 2\sqrt{2}}V_{cd}^*V_{cs}(c_1-c_2){2\sqrt{2}\over 3}(Y_1+9Y_2)(4\pi).
\en

In the bag model the form factors at the maximal recoil $q^2=0$ have the expressions \cite{Cheng:HFC}
\be
f_1^{\Lambda_b^0\Xi_b^-} =
f_1^{\Lambda_c^+\Xi_c^0} &=& \la \Lambda_c^+|b^\dagger_u b_s|\Xi_c^0\ra \int d^3{\bf r}(u_uu_s+v_uv_s)=4\pi Z_3, \non \\
f_1^{\Lambda_b^0\Xi_b^0} =
f_1^{\Lambda_c^+\Xi_c^+} &=& \la \Lambda_c^+|b^\dagger_d b_s|\Xi_c^+\ra \int d^3{\bf r}(u_du_s+v_dv_s)=-4\pi Z_3,
\en
and
\be
g_1^{\Xi_b^0\Omega_b^-}=
g_1^{\Xi_c^+\Omega_c^0} &=& \la \Xi_c^+|b^\dagger_u b_s\sigma_z|\Omega_c^0\ra \int d^3{\bf r}(u_uu_s-{1\over 3}v_uv_s)=-\sqrt{2\over 3}(4\pi Z_2),   \non \\
g_1^{\Xi_b^-\Omega_b^-}=
g_1^{\Xi_c^0\Omega_c^0} &=& \la \Xi_c^0|b^\dagger_d b_s\sigma_z|\Omega_c^0\ra \int d^3{\bf r}(u_du_s-{1\over 3}v_dv_s)=-\sqrt{2\over 3}(4\pi Z_2).
\en
The axial-vector form factor in the static limit can be expressed as
\begin{equation}
g^{A(P)}_{\mathcal{B}'\mathcal{B}}=\la \mathcal{B}' |b_{q_1}^\dagger b_{q_2}\sigma_z |
\mathcal{B}\ra \int d^3{\bf r}\left(u_{q_1}u_{q_2}-\frac13 v_{q_1}v_{q_2}\right).
\label{eq:gA}
\end{equation}
We find
\be
{1\over\sqrt{2}}g_{\Lambda_c^+\Sigma_c^0}^{A(\pi^-)}=g_{\Lambda_c^+\Sigma_c^+}^{A(\pi^0)}
=-g_{{\Xi'}_c^+\Xi_c^0}^{A(\pi^-)}=-g_{\Xi_c^+{\Xi'}_c^0}^{A(\pi^-)}=-2 g_{{\Xi'}_c^+\Xi_c^+}^{A(\pi^0)}=2 g_{\Xi_c^0{\Xi'}_c^0}^{A(\pi^0)}={1\over\sqrt{3}}(4\pi Z_1),
\en
with $Z_i$ being defined by
\be
&& Z_1=\int_0^R r^2dr\left(u_u^2-{1\over 3}v_u^2\right), \quad Z_2=\int_0^R r^2dr\left(u_uu_s-{1\over 3}v_uv_s\right),  \non  \\
&& \hskip 2 cm Z_3=\int_0^R r^2dr\left(u_uu_s+v_uv_s\right).
\en
Numerically, we obtain
\be
X_2=1.66\times 10^{-4}\,{\rm GeV}^3, \qquad
Y_1=8.37\times 10^{-6}\,{\rm GeV}^3, \qquad Y_2=2.11\times 10^{-4}\,{\rm GeV}^3,
\en
and
\be
 4\pi Z_1=0.65\,, \qquad 4\pi Z_2=0.71, \qquad 4\pi Z_3=0.985\,,
\en
where we have employed the following bag parameters
\be
m_u=m_d=0, \quad m_s=0.279~{\rm GeV}, \quad m_c=1.551~{\rm GeV}, \quad R=5~{\rm GeV}^{-1}.
\en

As shown in Ref. \cite{Cheng:HFC}, the matrix element of the 4-quark operator $O_1-O_2$ also can be evaluated in the so-called diquark model in which the matrix element $X$ defined in Eq. (\ref{eq:X&Y}) has the expression \cite{Cheng:HFC2016}
\be
X_{\rm di}={2\over 3\, m_{\rm di}} g_{du}g_{us},
\en
where $m_{\rm di}$ is the diquark mass taken to be 785 MeV and $g_{qq'}$ is the diquark decay constant. We shall follow \cite{diquark} to use
\be
(c_1-c_2)g_{du}g_{us}=0.066\pm0.013~{\rm GeV}^4,
\en
and hence
\be \label{eq:diquarkmodel}
X_{\rm di}={1\over c_1-c_2}\,(5.6\pm1.1)10^{-2}{\rm GeV}^3.
\en
Unfortunately, the diquark model is not applicable for estimating the baryon matrix elements of the operator $\tilde O_1-\tilde O_2$ such as $Y$ defined in Eq. (\ref{eq:X&Y}) because the heavy baryon is a bound state of the heavy quark and the light diquark in the heavy quark limit \cite{Cheng:HFC2016}.

There is an alternative approach for the evaluation of $Y$.
Voloshin \cite{Voloshin00,Voloshin14} has shown that, in the SU(3) limit, this baryon matrix element can be related to the matrix elements $x$ and $y$ defined by
\be \label{eq:x,y}
x\equiv -\la \Lambda_c^+|(\bar c\gamma_\mu c)(\bar d\gamma^\mu s)|\Xi_c^+\ra, \qquad
y\equiv -\la \Lambda_c^+|(\bar c_i\gamma_\mu c_k)(\bar d_k\gamma^\mu s_i)|\Xi_c^+\ra.
\en
These two parameters can be determined from the total decay width differences, namely, $\Gamma(\Xi_c^0)-\Gamma(\Lambda_c^+)$ and $\Gamma(\Lambda_c^+)-\Gamma(\Xi_c^+)$. For an updated estimate of $x$ and $y$, see \cite{Voloshin:2019}. Contrary to $B$ meson and bottom baryon cases where heavy quark expansion (HQE) in $1/m_b$ leads to the lifetime ratios in excellent agreement with experiment, HQE in $1/m_c$ does not work well for describing the lifetime pattern of charmed baryons. Since the charm quark is not heavy,
it is thus natural to consider the effects stemming from the next-order $1/m_c$ expansion.
It has been shown that $1/m_c$ corrections to spectator effects described by dimension-7 operators are very important for $\Gamma(\Lambda_c^+)$ and $\Gamma(\Xi_c^+)$ \cite{Cheng:2018}. In view of the recent new measurements of charmed baryon lifetimes and a new lifetime hierarchy $\tau(\Xi_c^+)>{\tau(\Omega_c^0)}>\tau(\Lambda_c^+)>\tau(\Xi_c^0)$ emerged (for a review, see Ref. \cite{Cheng:2021qpd}) the relation of $x$ and $y$ with the decay width differences will be much more subtle than before and has to be carefully re-examined.

The factorizable amplitudes are governed by the parameters $a_1$ and $a_2$ given in Eq. (\ref{eq:a1a2}). In the charmed baryon sector, $|a_2|$ has been determined from the decay $\Lambda_c^+\to p\,\phi$ to be $0.45\pm0.02$ \cite{CKX}. By treating $N_c$ in Eq. (\ref{eq:a1a2}) as an effective parameter, we have $N_c^{\rm eff}=18.8$ for $c_1=1.336$ and $c_2=-0.621$ at the $\mu=m_c$ scale. Indeed, it is well known that the large-$N_c$ approach works better in the charm sector. In bottom baryon decays, $|a_2|$ can be extracted from the branching fraction of $\Lambda_b^0\to J/\psi\Lambda$. However, the measured $\B(\Lambda_b^0\to J/\psi\Lambda)$ depends on the unknown branching fraction of $b\to \Lambda_b^0$. For our purpose, we shall take $a_2\approx 0.30$ as a benchmark. Then we have $N_c^{\rm eff}\approx 1.88$ for $c_1=1.139$ and $c_2=-0.307$ at the scale $\mu=\bar m_b(m_b)$.

Finally,  the decay rate and up-down spin asymmetry of $\B_i\to\B_f+P$ are given by
\be \label{eq:Gamma alpha}
\Gamma &=& {p_c\over 8\pi}\left\{ {(m_i+m_f)^2-m_P^2\over m_i^2} |A|^2+{(m_i-m_f)^2-m_P^2\over m_i^2} |B|^2\right\}, \non \\
 \alpha &=&\frac{2\kappa {\rm{Re}} (A^* B)}{|A|^2+\kappa^2|B|^2},
\en
with $p_c$ being the c.m. three-momentum in the rest frame of $\B_i$ and $\kappa=p_c/(E_f+m_f)=\sqrt{(E_f-m_f)/(E_f+m_f)}$.
For the heavy baryon lifetimes, we shall use \cite{PDG,LHCb:lifetime}
\be \label{eq:lifetimeWA}
&& \tau(\Xi_c^+)=(456\pm 5) ~{\rm fs}, \qquad\qquad~~ \tau(\Xi_c^0)=(152.0\pm 2.0) ~{\rm fs}, \qquad\quad~ \tau(\Omega_c^0)=(274.5\pm 12.4) ~{\rm fs},  \non \\
&& \tau(\Xi_b^0)=(1.480\pm0.030) ~{\rm ps}, \qquad \tau(\Xi_b^-)=(1.572\pm0.040) ~{\rm ps},
\qquad \tau(\Omega_b^-)=(1.64^{+0.18}_{-0.17}) ~{\rm ps}.  \non\\
\en

\section{Results and Discussion}

We first consider the decay $\Xi_b^-\to\Lambda_b^0\pi^-$ and compare our result with the LHCb measurement. From the calculations in Sec. IV we have
$f_1^{\Lambda_b^0\Xi_b^-}(m_\pi^2)\cong  f_1^{\Lambda_b^0\Xi_b^-}(0)=0.985$, $X=1.67\times 10^{-2}\,{\rm GeV}^3$ and $Y=5.51\times 10^{-3}\,{\rm GeV}^3$ in the bag model. This decay has only the $S$-wave amplitude and
it is evident from Eq. (\ref{eq:AmpHFC}) that the interference between factorizable and $W$-exchange amplitudes of $\Xi_b\to\Lambda_b\pi$ is destructive. The factorizable contribution was often neglected in the literature (see e.g. Refs. \cite{Voloshin00,Voloshin14}) because it is proportional to the pion's momentum which is small in HFC decays. Nevertheless, we will take into account the factorizable contribution in this work as it is not negligible. It turns out that the predicted $\B(\Xi_b^-\to\Lambda_b^0\pi^-)=6.6\times 10^{-4}$ is too small compared to the experimental range shown in Eq. (\ref{eq:BFXib LHCB}). This implies that the baryon matrix element $X$ defined in Eq. (\ref{eq:X&Y}) is underestimated in the bag model. In the diquark model $X_{\rm di}$ (see Eq. (\ref{eq:diquarkmodel}))  is larger than the bag model's estimate. The branching fraction of $\Xi_b^-\to\Lambda_b^0\pi^-$ then becomes $(4.7^{+2.3}_{-1.8})\times 10^{-3}$ which is consistent with the LHCb measurement given in Eq. (\ref{eq:BFXib LHCB}).
Since the matrix element $Y$ cannot be calculated in the diquark model, hereafter we will take the matrix elements in a hybrid manner: the diquark  model for $X$ and the bag model for $Y$.

We next consider the $S$-wave amplitudes of $\Xi_c^0\to\Lambda_c^+\pi^-$ and compare our results with that of Gronau and Rosner (GR) \cite{Gronau:HFC}. Since
$f_1^{\Lambda_c^+\Xi_c^0}=0.985$ we see from Eq. (\ref{eq:AmpHFCXic0}) that
the amplitudes of $A^{\rm fac}$ and  $A^{\rm nf}_{su\to  ud}$ are of the opposite sign and the interference between the $W$-exchange amplitudes induced from $cs\to dc$ and $su\to ud$ transitions is destructive. In Ref. \cite{Gronau:HFC}, GR denoted $A^{\rm fac}+A^{\rm nf}_{su\to ud}$ by $A_{s\to u\bar ud}$ and made the assumption that
\be
A_{s\to u\bar ud}(\Xi_c^0\to\Lambda_c^+\pi^-)=A(\Xi_b^-\to\Lambda_b^0\pi^-).
\en
The $S$-wave amplitude of $\Xi_b^-\to\Lambda_b^0\pi^-$ was then expressed in terms of two unknown parameters $F$ and $x$ which were in turn determined from a best fit to the observed nonleptonic hyperon decays \cite{Gronau:HFC}.
In our case, we performed a dynamic calculation of both $A^{\rm fac}$ and $A^{\rm nf}_{s u\to ud}$. The resultant value of $A_{s\to u\bar ud}$ is consistent with that of GR (see Table \ref{tab:comparisonofA}). It is the $W$-exchange term $A^{\rm nf}_{cs\to dc}$ that makes a huge difference. GR followed the method proposed by Voloshin \cite{Voloshin00,Voloshin14} to calculate the matrix element $Y$ and hence the amplitude $A^{\rm nf}_{cs\to dc}$ in terms of the matrix elements $x$ and $y$ introduced in Eq. (\ref{eq:x,y}). We have already remarked in Sec. IV that the issue of determining $x$ and $y$ in terms of the decay width differences should be re-examined in view of important $1/m_c$ corrections to $\Gamma(\Lambda_c^+)$ and $\Gamma(\Xi_c^+)$
described by dimension-7 operators in HQE. The estimate of $A^{\rm nf}_{cs\to dc}$ by GR is shown in Table \ref{tab:comparisonofA}. As the relative sign between $A(\Xi_b^-\to\Lambda_b^0\pi^-)$ and
$A^{\rm nf}_{cs\to dc}$ is not fixed, GR considered both constructive and destructive interference. In our dynamic calculation, it is evident from Eq. (\ref{eq:AmpHFCXic0}) that the interference is destructive because of the relation $V_{cd}^*V_{cs}= -V_{ud}^*V_{us}$ for CKM matrix elements. In the case of destructive interference, both this work and GR have similar branching fractions of order $10^{-4}$ from the $S$-wave alone,

\begin{table}[t]
\begin{ruledtabular}
 \caption{$S$-wave amplitudes (in units of $10^{-7}$) of $\Xi_c^0\to\Lambda_c^+\pi^-$ and the branching fractions calculated in this work and Ref. \cite{Gronau:HFC}. Destructive and constructive interferences between $A^{\rm fac}+A^{\rm nf}_{su\to ud}$ and  $A^{\rm nf}_{sc\to ud}$ amplitudes shown in the first and second entries, respectively, were both considered in \cite{Gronau:HFC}, in which the results have been normalized using the current world averages of charmed baryon lifetimes  (see Eq. (\ref{eq:lifetimeWA})).
 } \label{tab:comparisonofA}
\vspace{6pt}
\begin{tabular}
{ l c c c c  }
   & $A^{\rm fac}+A^{\rm nf}_{su\to ud}$ & $A^{\rm nf}_{sc\to cd}$ & $A^{\rm tot}$ & $\B_{\rm S-wave}$  \\
 \hline
This work & $3.27\pm0.75$ \footnotemark[1]
 & $-0.74$ &  $2.53\pm0.75$  & $(2.53^{+1.74}_{-1.28})\times 10^{-4}$    \\
 Gronau, Rosner \cite{Gronau:HFC} &   $3.97\pm0.59$ & $-1.86\pm0.91$ &  $2.11\pm1.08$  & $(1.76^{+2.26}_{-1.34})\times 10^{-4}$  \\
 & $3.97\pm0.59$
 & ~\,$1.86\pm0.91$ &  $5.83\pm1.08$  & $(1.34\pm0.53)\times 10^{-3}$  \\
\end{tabular}
\end{ruledtabular}
\footnotetext[1]{Explicitly, $A^{\rm fac}=-0.56$ and $A^{\rm nf}_{su\to ud}=3.83\pm0.75$ in unit of $10^{-7}$.}
\end{table}

\begin{table}[t]
\begin{ruledtabular}
 \caption{The magnitude of $S$- and $P$-wave amplitudes (in units of $10^{-7}$), branching fractions and decay asymmetries of the HFC decays of charmed and bottom baryons. Uncertainties arise from the matrix element $X$ estimated in the diquark  model, see Eq. (\ref{eq:diquarkmodel}).
 Experimental measurements are taken from Refs. \cite{LHCb:HFCb,LHCb:HFC}.
 } \label{tab:HFC:AmpandBF}
 \vspace{6pt}
\begin{tabular}
{ l c c c c  }
   & $\Xi_c^0\to \Lambda_c^+\pi^-$ & $\Xi_c^+\to \Lambda_c^+\pi^0$ & $\Xi_b^-\to \Lambda_b^0\pi^-$ & $\Xi_b^0\to \Lambda_b^0\pi^0$ \\
 \hline
 $A$ & $2.53\pm 0.75$ & $2.02\pm0.53$ & $3.43\pm0.76$ & $2.80\pm0.53$ \\
 $B$ & $\!\!244$ & $\!\!181$ & 0 & 0 \\
 \hline
  $\alpha$ & $0.70^{+0.13}_{-0.17}$ & $0.74^{+0.11}_{-0.16}$ & 0 & 0 \\
 $\B$ & $(1.76^{+0.18}_{-0.12})\times 10^{-3}$ & $(3.03^{+0.29}_{-0.22})\times 10^{-3}$  & $(4.67^{+2.29}_{-1.83})\times 10^{-3}$ & $(2.87^{+1.20}_{-0.99})\times 10^{-3}$  \\
  $\B_{\rm expt}$ & $(5.5\pm1.8)\times 10^{-3}$ & -- & see Eq. (\ref{eq:BFXib LHCB})  & -- \\
\end{tabular}
\end{ruledtabular}
\end{table}

Although the obtained $S$-wave is too small to account for the observed rate for the HFC decay of $\Xi_c^0$, the story on the PC part is entirely unexpected. Very recently,
Niu, Wang and Zhao (NWZ) \cite{Niu:2021qcc}, Groote and K\"orner (GK) \cite{Groote:2021pxt} have independently pointed out the importance of the $\Sigma_c$ pole in the consideration of nonspectator $W$-exchange contribution to the $P$-wave amplitude (see Eq. (\ref{eq:Pwavenet})).
The calculated $S$- and $P$-wave amplitudes and branching fractions are displayed
in Table \ref{tab:HFC:AmpandBF}. Evidently, $\Xi_c\to\Lambda_c^+\pi$ decays are dominated by the PC transition governed mainly by the $\Sigma_c$ intermediate state. Numerically, the $P$-wave is larger than the $S$-wave by two orders of magnitude, in agreement with NWZ. Although both PC and PV amplitudes of  $\Xi_c^0\to \Lambda_c^+\pi^-$ are larger than that of $\Xi_c^+\to \Lambda_c^+\pi^0$, the branching fraction of the former is smaller than the latter because the lifetime of $\Xi_c^+$ is three times longer than that of $\Xi_c^0$. Contrary to the charmed baryon sector, HFC decays of $\Xi_b$ proceed through the PV transition and the $b$ quark acts as a spectator.

In Table \ref{tab:HFC:AmpandBF} we also show the numerical results for the asymmetry parameter $\alpha$. It is zero for $\Xi_b\to\Lambda_b\pi$ due to the absence of PC amplitudes. Naively, it was argued in Ref. \cite{Niu:2021qcc} that $\alpha$ is very small in $\Xi_c\to \Lambda_c\pi$ because the magnitude of the $P$-wave is larger than that of the $S$-wave by two orders of magnitude. This is not the case in reality as the parameter $\kappa$ introduced in Eq. (\ref{eq:Gamma alpha}) is very small, of order $10^{-2}$, owing to the soft momentum of the pion. Consequently, we find that $\alpha$ is {\it positive} and of order $0.70$ and $0.74$ for
$\Xi_c^0\to \Lambda_c^+\pi^-$ and $\Xi_c^+\to \Lambda_c^+\pi^0$, respectively. In the work of NWZ, $S$- and $P$-wave amplitudes are of opposite sign (see Table VII of Ref. \cite{Niu:2021qcc}). Hence, a measurement of the asymmetry parameter will enable to discern different model predictions.

In Table \ref{tab:ComparisonBR} we compare our results with other model calculations for HFC $\Xi_c\to\Lambda_c^+\pi$ decays. It is clear that all the previous predictions excluding $P$-wave contributions are too small compared to experiment. We have noticed in passing that the interference between $W$-exchange amplitudes through $cs\to dc$ and $su\to ud$ transition should be destructive. This will render the branching fractions for $\Xi_c\to\Lambda_c^+\pi$ even smaller. The observed sizable $\B(\Xi_c^0\to\Lambda_c^+\pi^-)$ strongly indicates the importance of PC amplitudes. The calculated result for this mode by NWZ is consistent with the LHCb measurement. \footnote{Here we make one remark on the work of NWZ \cite{Niu:2021qcc}. It appears that factorizable internal $W$-emission contributions were not considered by the authors. The color-suppressed process depicted in Fig. 1(b) of Ref. \cite{Niu:2021qcc} is not a factorizable diagram. To have an internal $W$-emission contribution, the emitted meson must be formed from $u_1$ and $\bar u_4$ induced from the $s$ quark decay and the middle $s$ quark line should be put on top of the $q$ quark line.  Indeed, it is easily seen that Fig. 1(b) is equivalent to Fig. 2(d); it has the same topology as the $W$-exchange diagram. It is known that $\Xi_c^+\to \Lambda_c^+\pi^0$ has a factorizable internal $W$-emission contribution, while  $\Xi_c^0\to \Lambda_c^+\pi^-$ receives an external $W$-emission term. In Table VII of Ref. \cite{Niu:2021qcc}, we see that both modes receive contributions from the ``CS" diagram.
This also indicates that the ``CS" term is actually not referred to the color-suppressed factorizable contribution.
}
In our work, the predicted value of $\B(\Xi_c^0\to\Lambda_c^+\pi^-)$ is slightly smaller than experiment. It is not unexpected as we have noticed in passing that the matrix element $X$ is underestimated in the bag model. It is most likely that the matrix element $Y$ is also underestimated in the same model. Obviously,
we need to improve the estimate the matrix elements of the 4-quark operators $(\bar dc)(\bar ss)$ and $(\bar cc)(\bar ds)$.
Another source of theoretical uncertainty stems from the Wilson coefficients $c_1$ and $c_2$ as the baryon matrix element depends on the Wilson coefficient difference $c_1-c_2$.

\begin{table}[tp!]
\begin{ruledtabular}
 \caption{Branching fractions (in units of $10^{-3}$) of charm-flavor-conserving decays $\Xi_c\to\Lambda_c^+\pi$. All the model results
 have been normalized using the current world averages of lifetimes for $\Xi_c^+$ and $\Xi_c^0$ (see Eq. (\ref{eq:lifetimeWA})). As for the predictions of \cite{Gronau:HFC}, the first (second) entry is for destructive (constructive) interference between $A^{\rm fac}+A^{\rm nf}_{su\to ud}$ and  $A^{\rm nf}_{sc\to ud}$ amplitudes.
 } \label{tab:ComparisonBR}
\vspace{6pt}
\footnotesize{
\begin{tabular}
{ l c c c c c c c c}
 Mode  & (CLY)$^2_a$ & (CLY)$^2_b$ & Faller & Gronau & Voloshin & Niu & This work & Experiment \\
 & \cite{Cheng:HFC} & \cite{Cheng:HFC2016} & \cite{Faller} & \cite{Gronau:HFC} & \cite{Voloshin:2019} &  \cite{Niu:2021qcc} & & \cite{LHCb:HFC} \\
 \hline
$\Xi_{c}^{0}\to\Lambda_c^+\pi^-$ & $0.39$   & $0.17$  & $<3.9$ & $0.18^{+0.23}_{-0.13}$ & $>\!0.25\pm0.15$  & $5.8\pm2.1$ & $1.76^{+0.18}_{-0.12}$  & $5.5\pm0.2\pm1.8$  \\
&    &   &  & $1.34\pm0.53$ &   &  &  &   \\
$\Xi_{c}^{+}\to\Lambda_c^+\pi^0$ & $0.69$   & $0.11$  & $<6.1$ & $<\!0.2$ & -- & $11.1\pm4.0$ & $3.03^{+0.29}_{-0.22}$ & --\\
&  &   & & $2.01\pm0.80$ & & & \\
\end{tabular}
}
 \end{ruledtabular}
 \end{table}

\begin{table}[t]
\begin{center}
 \caption{The magnitude of $S$- and $P$-wave amplitudes (in units of $10^{-7}$), branching fractions and decay asymmetries of the HFC decays $\Omega_c^0\to \Xi_c\pi$ and $\Omega_b^-\to\Xi_b \pi$ calculated using the bag model.
 } \label{tab:HFC_OmegaQ}
 \vspace{6pt}
\begin{tabular}
{ l c c c c  } \hline \hline
   & ~~~$\Omega_c^0\to \Xi_c^+\pi^-$~~~ & ~~~$\Omega_c^0\to \Xi_c^0\pi^0$~~~ & ~~~$\Omega_b^-\to \Xi_b^0\pi^-$~~~ & ~~~$\Omega_b^-\to \Xi_b^-\pi^0$~~~ \\
 \hline
 $A$ & $-1.72$ & ~~~~$1.21$ & 0 & 0 \\
 $B$ & $40.12$ & $-32.19$ & $-15.96$ & $-3.48$ \\
 \hline
 $\B$ & $5.1\times 10^{-4}$ & $2.8\times 10^{-4}$  & $6.5\times 10^{-5}$  & $2.8\times 10^{-6}$  \\
 $\alpha$ & $-0.98$ & $-0.99$ & 0 & 0 \\
 \hline \hline
\end{tabular}
\end{center}
\end{table}

As for HFC decays $\Omega_Q\to \Xi_Q\pi$, the numerical results are exhibited in Table \ref{tab:HFC_OmegaQ}, where uses of Eqs. (\ref{eq:SwaveOMegac}) and (\ref{eq:PwaveOmegaQ}) as well as model calculations of various baryon matrix elements and axial-vector form factors in Sec. IV have been made. The decays $\Omega_b\to \Xi_b\pi$ receive only factorizable $P$-wave contributions, while $\Omega_c\to \Xi_c\pi$ acquire additional contributions from nonspectator
$W$-exchange for both PC and PV amplitudes, \footnote{$W$-exchange contributions to $\Omega_c\to\Xi_c\pi$ were also discussed recently in Ref. \cite{Groote:2021pxt}. Apart from the signs which depend on the convention for wave functions, the obtained relations (see Table 15 of Ref. \cite{Groote:2021pxt})
\be
2\sqrt{6}\,a_{\Xi_c^0\Omega_c^0}= 2\sqrt{2}\,a_{{\Xi'}_c^0\Omega_c^0}=-4\sqrt{3}\,a_{\Lambda_c^+{\Xi'}_c^+}
=-2\sqrt{6}\,a_{\Sigma_c^0\Xi_c^0}
=12\la\Lambda_c^+|{\cal H}^{(c)}_{\rm eff}|\Xi_c^+\ra   \non
\en
among various matrix elements relevant for HFC decays of charmed baryons agree with our Eqs. (\ref{eq:XY}), (\ref{eq:m.e.Xic}) and (\ref{eq:meofOmegaQ}) provided that the bag integral $Y_1$, which is smaller than $Y_2$, is neglected.
}
which satisfy the $\Delta I=1/2$ relation
\be
M(\Omega_c^0\to \Xi_c^+\pi^-)=-\sqrt{2}M(\Omega_c^0\to \Xi_c^0\pi^0),
\en
for both PC and PV transitions. Since the mass difference between $\Omega_c^0$ and the intermediate ${\Xi'}_c^0$ pole
is 116 MeV, the $P$-wave enhancement is not so dramatic as in the case of $\Xi_c\to \Lambda_c\pi$. The predicted branching fraction is of order $5\times 10^{-4}$ for $\Omega_c^0\to\Xi_c^+\pi^-$ and $3\times 10^{-4}$ for $\Omega_c^0\to\Xi_c^0\pi^0$ and smaller for $\Omega_b\to\Xi_b\pi$. The asymmetry parameter $\alpha$ vanishes in the decays $\Omega_b\to\Xi_b\pi$ owing to the absence of the $S$-wave transition. By contrast, it is close to $-1$ in $\Omega_c\to\Xi_c\pi$ modes.

\section{Conclusion}
In this work we have studied HFC hadronic weak decays of heavy baryons.
It was pointed out three decades ago that if the heavy quark in the HFC process acts as a spectator, the $P$-wave amplitude of the $\B_{\bar 3}\to \B_{\bar 3}+P$ decay such as $\Xi_Q\to\Lambda_Q\pi$ will vanish in the heavy quark limit. Indeed, this is the case for $\Xi_b\to\Lambda_b \pi$ decays. For $\Xi_c\to\Lambda_c\pi$ decays, they receive  additional nonspectator contributions arising from the $W$-exchange diagrams through the $cs\to dc$ transition. Our main results are:

\begin{itemize}
\item
We have shown explicitly that the factorizable  and nonfactorizable $P$-wave amplitudes of $\Xi_Q\to \Lambda_Q\pi$ vanish because of vanishing $\B_{\bar 3}\B_{\bar 3}\pi$ strong coupling and the absence of $\B_{\bar 3}-\B_6$ weak transition in the limit of heavy quark symmetry, provided that the heavy quark does not participate in weak interacions.

\item
In the presence of the nonspectator $W$-exchange via $cs\to dc$, it contributes to the $S$-wave destructively to render the PV amplitude even smaller.

\item
If the baryon matrix elements of 4 light-quark operators denoted by $X$ are evaluated in the bag model, the predicted branching fraction of $\Xi_b^-\to \Lambda_b^0\pi^-$ will be too small compared to experiment. Hence, we use the diquark model to estimate the baryon matrix element $X$. Nevertheless, we still use the bag model to evaluate the baryon matrix element of 4-quark oprtators involving $c$ and $\bar c$ as the diquark model is not applicable in this case.

\item
We have confirmed that the HFC decays  $\Xi_c\to\Lambda_c\pi$ are dominated by the PC pole terms induced from the nonspectator $W$-exchange.
The small mass difference between $\Xi_c$ and the intermediate $\Sigma_c$ pole, of order 16 MeV, leads to a strong enhancement of the $P$-wave pole amplitudes.

\item
Contrary to the HFC decays $\Xi_b\to\Lambda_c\pi$, $\Omega_b\to \Xi_b\pi$ modes receive only factorizable $P$-wave contributions. In the charm sector, $\Omega_c\to \Xi_c\pi$ acquire additional contributions from nonspectator $W$-exchange for both PC and PV amplitudes. The $P$-wave transition is enhanced by the ${\Xi'}_c$ pole, though it is not so dramatic as in the case of $\Xi_c\to\Lambda_c\pi$. The predicted branching fraction is of order $5\times 10^{-4}$ for $\Omega_c^0\to\Xi_c^+\pi^-$ and $3\times 10^{-4}$ for $\Omega_c^0\to\Xi_c^0\pi^0$.

\item
The asymmetry parameter $\alpha$ vanishes in the decays $\Xi_b\to \Lambda_b\pi$ and $\Omega_b\to\Xi_b\pi$ owing to the absence of $P$- and $S$-wave transitions, respectively. By contrast, it is close to $-1$ in $\Omega_c\to\Xi_c\pi$ modes. For decays $\Xi_c^0\to\Lambda_c^+\pi^-$ and $\Xi_c^+\to\Lambda_c^+\pi^0$, the decay asymmetries are found to be positive, of order $0.70$ and $0.74$, respectively.

\item
The major theoretical uncertainty arises from the evaluation of the baryon matrix elements of 4-quark operators, which needs to be improved in the future.

\end{itemize}

\vskip 2 cm

\section*{Acknowledgments}

We are grateful to Dr. Peng-Yu Niu for very helpful discussions. This research was supported in part by the Ministry of Science and Technology of R.O.C. under Grant No. MOST-110-2112-M-001-025,  the National Natural Science Foundation of China
under Grant Nos. U1932104 and 12142502, and the Guangdong
Provincial Key Laboratory of Nuclear Science with No. 2019B121203010.

\newcommand{\bi}{\bibitem}


\begin{thebibliography}{99}

\bibitem{Cheng:HFC}
H.~Y.~Cheng, C.~Y.~Cheung, G.~L.~Lin, Y.~C.~Lin, T.~M.~Yan and H.~L.~Yu,
  ``Heavy flavor conserving nonleptonic weak decays of heavy baryons'',
  Phys.\ Rev.\ D {\bf 46}, 5060 (1992).

\bibitem{Yan}
  T.~M.~Yan, H.~Y.~Cheng, C.~Y.~Cheung, G.~L.~Lin, Y.~C.~Lin and H.~L.~Yu,
  ``Heavy quark symmetry and chiral dynamics'',
  Phys.\ Rev.\ D {\bf 46}, 1148 (1992)
  [Phys.\ Rev.\ D {\bf 55}, 5851(E) (1997)].

\bibitem{Wise} M. B. Wise,
``Chiral perturbation theory for hadrons containing a heavy quark",
Phys. Rev. D {\bf 45}, 2188 (1992);
G. Burdman and J. Donoghue,
``Union of chiral and heavy quark symmetries",
Phys. Lett. B {\bf 280}, 287 (1992).

\bibitem{Sinha:HFC}
S.~Sinha and M.~P.~Khanna,
``Beauty-conserving strangeness-changing two-body hadronic decays of beauty baryons,''
Mod. Phys. Lett. A \textbf{14}, 651-660 (1999)

\bibitem{Voloshin00}  M.~B.~Voloshin,
  ``Weak decays $\Xi_Q \to \Lambda_Q \pi$,"
  Phys.\ Lett.\ B {\bf 476}, 297 (2000)
  [hep-ph/0001057].

\bibitem{Voloshin14}
  X.~Li and M.~B.~Voloshin,
  ``Decays $\Xi_b \to \Lambda_{b} \pi$ and diquark correlations in hyperons,"
  Phys.\ Rev.\ D {\bf 90},  033016 (2014)
  [arXiv:1407.2556 [hep-ph]].


\bibitem{Faller}
S.~Faller and T.~Mannel,
``Light-Quark Decays in Heavy Hadrons,''
Phys. Lett. B \textbf{750}, 653-659 (2015)
[arXiv:1503.06088 [hep-ph]].

\bibitem{Cheng:HFC2016}
H.~Y.~Cheng, C.~Y.~Cheung, G.~L.~Lin, Y.~C.~Lin, T.~M.~Yan and H.~L.~Yu,
``Heavy-Flavor-Conserving Hadronic Weak Decays of Heavy Baryons,''
JHEP \textbf{03}, 028 (2016)
[arXiv:1512.01276 [hep-ph]].

\bibitem{Gronau:2015jgh}
M.~Gronau and J.~L.~Rosner,
``$S$-wave nonleptonic hyperon decays and $\Xi^-_b \to \pi^- \Lambda_b$,''
Phys. Rev. D \textbf{93}, no.3, 034020 (2016)
[arXiv:1512.06700 [hep-ph]].


\bibitem{Gronau:HFC}
M.~Gronau and J.~L.~Rosner,
``From $\Xi_b \to \Lambda_b \pi$ to $\Xi_c \to \Lambda_c \pi$,''
Phys. Lett. B \textbf{757}, 330-333 (2016)
[arXiv:1603.07309 [hep-ph]].


\bibitem{Voloshin:2019}
M.~B.~Voloshin,
``Update on splitting of lifetimes of $c$ and $b$ hyperons within the heavy quark expansion and decays $\Xi_Q \to \Lambda_Q \pi$,''
Phys. Rev. D \textbf{100}, no.11, 114030 (2019)
[arXiv:1911.05730 [hep-ph]].

\bibitem{Niu:2021qcc}
P.~Y.~Niu, Q.~Wang and Q.~Zhao,
``Study of heavy quark conserving weak decays in the quark model,''
Phys. Lett. B \textbf{826}, 136916 (2022)
[arXiv:2111.14111 [hep-ph]].

\bibitem{Groote:2021pxt}
S.~Groote and J.~G.~K\"orner,
``Topological tensor invariants and the current algebra approach: Analysis of 196 nonleptonic two-body decays of single and double charm baryons - a review,''
Eur. Phys. J. C \textbf{82} (2022) 297
[arXiv:2112.14599 [hep-ph]].

\bibitem{LHCb:HFCb}
R.~Aaij \textit{et al.} [LHCb Collaboration],
``Evidence for the strangeness-changing weak decay $\Xi_b^-\to\Lambda_b^0\pi^-$,''
Phys. Rev. Lett. \textbf{115}, no.24, 241801 (2015)
[arXiv:1510.03829 [hep-ex]].

\bibitem{LHCb:HFC}
R.~Aaij \textit{et al.} [LHCb Collaboration],
``First branching fraction measurement of the suppressed decay $\Xi_c^0\to \pi^-\Lambda_c^+$,''
Phys. Rev. D \textbf{102}, no.7, 071101 (2020)
[arXiv:2007.12096 [hep-ex]].

\bibitem{Buchalla}
  G.~Buchalla, A.~J.~Buras and M.~E.~Lautenbacher,
  ``Weak decays beyond leading logarithms'',
  Rev.\ Mod.\ Phys.\  {\bf 68}, 1125 (1996)
  [hep-ph/9512380].

\bibitem{PDG}
P.A. Zyla {\it et al.} [Particle Data Group], Prog. Theor. Exp. Phys. 2020, 083C01 (2020) and 2021 update.

\bibitem{Cheng:1992}
  H.~Y.~Cheng and B.~Tseng,
  ``Nonleptonic weak decays of charmed baryons'',
  Phys.\ Rev.\ D {\bf 46}, 1042 (1992).
  [Phys.\ Rev.\ D {\bf 55}, 1697(E) (1997)].

\bibitem{CKX}
  H.~Y.~Cheng, X.~W.~Kang and F.~Xu,
  ``Singly Cabibbo-suppressed hadronic decays of $\Lambda_c^+$,''
  Phys.\ Rev.\ D {\bf 97}, 074028 (2018)
  [arXiv:1801.08625 [hep-ph]].

\bibitem{Zou:2019kzq}
J.~Zou, F.~Xu, G.~Meng and H.~Y.~Cheng,
``Two-body hadronic weak decays of antitriplet charmed baryons,''
Phys. Rev. D \textbf{101}, no.1, 014011 (2020)
[arXiv:1910.13626 [hep-ph]].

\bibitem{MIT}
  A.~Chodos, R.~L.~Jaffe, K.~Johnson and C.~B.~Thorn,
  ``Baryon Structure in the Bag Theory'',
  Phys.\ Rev.\ D {\bf 10}, 2599 (1974);
  T.~A.~DeGrand, R.~L.~Jaffe, K.~Johnson and J.~E.~Kiskis,
  ``Masses and Other Parameters of the Light Hadrons'',
  Phys.\ Rev.\ D {\bf 12}, 2060 (1975).


\bibitem{diquark}
  M.~Jamin and M.~Neubert,
  ``Diquark Decay Constants From {QCD} Sum Rules'',
  Phys.\ Lett.\ B {\bf 238}, 387 (1990);
  M.~Neubert and B.~Stech,
  ``A Consistent analysis of the $\Delta I = 1/2$ rule in strange particle
  decays'',
  Phys.\ Rev.\ D {\bf 44}, 775 (1991).

\bibitem{Cheng:2018}
 H.~Y.~Cheng,
  ``Phenomenological Study of Heavy Hadron Lifetimes,''
  JHEP {\bf 1811}, 014 (2018)
  [arXiv:1807.00916 [hep-ph]].

\bibitem{Cheng:2021qpd}
H.~Y.~Cheng,
``Charmed Baryon Physics Circa 2021,''
[arXiv:2109.01216 [hep-ph]].


\bibitem{LHCb:lifetime}
R.~Aaij \textit{et al.} [LHCb Collaboration],
``Measurement of the lifetimes of promptly produced $\Omega^{0}_{c}$ and $\Xi^{0}_{c}$ baryons,''
Sci. Bull. \textbf{67}, 479-487 (2022)
[arXiv:2109.01334 [hep-ex]].



\end{thebibliography}
\end{document}